\documentclass{aa}
\usepackage{graphicx}

\begin {document}

\title {Catalogue of Apparent Diameters and Absolute Radii of Stars (CADARS) --
Third Edition -- Comments and Statistics}

\author{L.E.~Pasinetti~Fracassini\inst{1} \and
L.~Pastori\inst{2} \and
S.~Covino\inst{2} \and
A.~Pozzi\inst{3}}

\offprints{L.E.~Pasinetti~Fracassini}

\institute {Dipartimento di Fisica, Universit\'a degli Studi, Via G.Celoria 16, 
I-20133 Milano, Italy \and
Osservatorio Astronomico di Brera, Via Bianchi 46, I-23807 
Merate, Italy \and Thesis for the Degree in Physics}

\date {Received date; Accepted Date}

\abstract{
The Catalogue, available at the Centre de Donn\'ees
Stellaires de Strasbourg, consists of 13573 records concerning the
results obtained from different methods for 7778 stars, reported in
the literature. The following data are listed for each star: identifications,
apparent magnitude, spectral type, apparent diameter in arcsec, 
absolute radius in solar units, method of determination, reference,
remarks. Comments and statistics obtained from CADARS are given.
}

\titlerunning{Catalogue of Apparent Diameters...}
\authorrunning{L.E.~Pasinetti et al.}

\maketitle

\section{Introduction}

The first edition of the CADARS reporting data published up to 1979, appeared in 
1981 (Fracassini \& Pasinetti 1979; Fracassini et al. 1981b), and was installed 
in the CDS catalogue service as cat. II/61. Preliminary comments were reported 
by Fracassini et al. (1981a). Since then, a large amount of data appeared in the 
literature, many of them obtained from the modern interferometric techniques. 
The second edition, completely revised and updated, was CDS catalogue II/155 
(Fracassini et al. 1988, Pastori et al. 1988) and was reported in ``Selected 
Astronomical Catalogues", Vol.\,2, CD-ROM, ADC, NASA. Third edition, installed 
as 
CDS cat. II/224, consists of 13573 records, more than twice the number of the 
first edition. Actually, the number of available data is higher as for many 
stars both the apparent diameter and absolute radius are given. Moreover, data 
obtained by the same author in different wavelengths are also given in the 
remarks. The records concern the results obtained from different methods 
for 7778 stars, including stars of the Magellanic Clouds and two neutron stars.

\section{Classification of the methods} 

The methods for the determination of stellar dimensions were classified as 
direct or indirect methods. The direct methods are based on the observation
of some physical phenomena {\it directly} correlated with the geometry of the 
stellar disks. The indirect methods are based on the observation of some 
physical parameters {\it indirectly} correlated with the geometry of the 
stellar disks. A more detailed subdivision is based on the physical 
principles of determination (Fracassini et al. 1981a). A recent discussion on 
the determination methods is given by Scholz (1997). Table\,\ref{tab:1} 
summarizes the classification adopted in the first edition of CADARS 
(Fracassini et al. 1981b) and in the updates. Column 1 reports the code number 
given in the Catalogue 
to each method of determination, column 2 the corresponding method, and 
eventually column 3 the reference of the first measures and/or {\it basic} 
paper. An adopted criterion was to restrict as much as possible the number of 
codes and to include in the same group all the methods based on similar 
fundamental principles. Therefore, only one code has been added in the third 
edition.

\begin{table*}
\begin{center}
{\vskip 0.25cm \bf \large DIRECT METHODS \vskip 0.25cm}
\begin{tabular}{ccc}
\hline
CODE  &       METHOD          &    REFERENCES  \\
\hline
 1  & Interferometer           &  {\small Michelson A.A., Pease F.G. 1921, ApJ 53, 249}\\  
    & Intensity interferometer &  {\small Hanbury Brown R., Davis J. et al. 1967, MNRAS 
137, 393}\\
 2  & Diffraction              &  {\small Arnulf A. 1936, C.R.Acad.Sci.Paris 202, 115}\\
    & Lunar occultations       &  {\small Williams J.D. 1939, ApJ 89, 467}\\
 3A & Star scintillation -- Color changes &  {\small Tichov G. 1921, Mitt.Leshafts 
Inst.Leningrad 2, 126}\\
 3B & Star scintillation -- speckle interferometry   &  {\small Gezari D.Y., Labeyrie 
A., Stachnik R.V. 1972, ApJ 173, L1}\\
 4  & Eclipsing and spectroscopic binaries &  {\small Russell H.N. 1911, ApJ 35, 315}\\ 
    &                                      & {\small Lehman-Filh\'es R. 1894, Astron.Nach. 130, 17}\\
 5  & Pulsating stars          &  {\small Van Hoof A. 1945, Publ.Lab.Astron. 
G\'eod\'es.,Univ.Louvain XI, 100} \\
 5A & Pulsating stars          &  {\small Wesselink A.J. 1946, Bull.Inst.Netherlands X, 
91}\\
 5B & Pulsating stars          &  {\small Balona L.A. 1977, MNRAS 178, 231}\\
 5C & Pulsating stars          &  {\small Methods which cannot be included in the groups 
5, 5A, 5B}\\
\hline
\end{tabular}
{\vskip 0.25cm \bf \large INDIRECT METHODS \vskip 0.25cm} 
\begin{tabular}{ccc}
\hline
CODE  &       METHOD          &    REFERENCES  \\
\hline
 6A & Intrinsic brightness and color &   Pickering E.C. 1880, 
Proc.Amer.Acad.Arts \&
Sci. 16, 1\\
 6B &      ``         "         &   {\small Russell H.N. 1920, PASP 32, 307}\\
 6C &      ``         "         &   {\small Hertzsprung E. 1922, Ann.Leiden XIV, 1}\\
 6D &      ``         "         &   {\small Pettit E., Nicholson S. 1928, ApJ 68, 279}\\
 6E &      ``         "         &   {\small Chalonge D., Divan L. 1950, 
C.R.Acad.Sci.Paris 231, 331}\\
 6F &      ``         "         &   {\small Fracassini M., Pasinetti L.E. 1967, Atti XI 
Riunione SAIt. Padova} \\
    &      ``         "         &   {\small Fracassini M., Gilardoni G., Pasinetti 
L.E.1973, Ap\&SS 22, 141}\\
 6G &      ``         "        &   {\small Gray D.F. 1967, ApJ 149, 317}\\
    &      ``         "        &   {\small Blackwell D.E., Shallis M.J. 1977, MNRAS 180, 
177}\\
 6H &      ``         "        &    {\small Wesselink A.J. 1969, MNRAS 144, 297}\\
 6I &      ``         "        &    {\small Barnes T.G., Evans D.S. 1976, MNRAS 
174, 489}\\
 6K &      ``         "        &    {\small Leone S. 1978, Atti Acc.Sci.Lettere, Arti 
Palermo, Ser.IV, 35, 21}\\
 6L &      ``         "        &    {\small Walker H.J., Sch\"onberner D. 1981, A\&A 97, 
291}\\ 
 6M &  Fundamental stellar parameters &   {\small Various authors}\\
\hline
\end{tabular}
\end{center}
\caption{Classification of the methods and codes} 
\label{tab:1}
\end{table*}

\section{The catalogue}

CADARS reports data appeared in the literature since 1950. However, fundamental 
and/or interesting data obtained before this year were also reported according 
to criteria given by Fracassini et al. (1981b). Third edition includes measures 
published from 1986 to 1997. 

The stars are listed according to the following order of identification: HD 
number, DM number, variables with constellation name in alphabetical order of 
the abbreviations, other identifications in alphabetical order, LMC and SMC 
stars, neutron stars at the end of CADARS (included only in the case of data 
derived from more or less direct measures). For the nomenclature of the 
stars see
Fracassini et al. (1981b, Table II) and/or SIMBAD. At least one identification 
is that from the author. The catalogue is followed by the lists of the remarks 
and references. The columns of CADARS are the following:
\begin{enumerate}
\item Identifier: HD number (first priority); DM number (secondary priority); 
constellation name in alphabetical order (variable stars), when HD or DM numbers 
are not available; other identifications.
\item Bayer or Flamsteed designation or other identification; G and S indicate 
the components of a binary system.
\item Apparent visual magnitude given by the authors, otherwise by SIMBAD; 
variable stars: magnitude at maximum luminosity; remarks (W) specify magnitudes 
in other wavelengths.
\item Spectral type and luminosity class given by the authors, otherwise by 
SIMBAD. \item Apparent diameter (arcsec) of the uniform disk; remarks specify 
other cases and/or values corrected for the limb darkening (L).Three significant 
figures are given; values given with more figures are reported in remarks. 
Errors (E) are given in remarks for uniformity with the previous versions.
\item Absolute radius (solar units). Three significant figures are given. Errors 
in remarks. For the Cepheids (code 5, 5A, 5B): averaged absolute radius given
by the authors.
\item Method of determination (see Table 1 for the codes).
\item Source by an alphanumerical code. 1st digit: initial of the first author. 
2nd digit: progressive number (see references).
\item Remarks. In the file of the remarks the star identifier is followed by the 
code of the source. Remark codes:
\begin{enumerate} 
\item B Spectroscopic binary
\item D Double star
\item G Cluster, group, aggregate or association membership;
\item R Value of the apparent diameter and/or absolute radius corrected 
for the interstellar 
reddening. This remark is not used for the values from method 6F, all corrected 
for the reddening.
\end{enumerate}
\item Codes calling attention to the file of the remarks:
\begin{enumerate}
\item A General characteristics
\item E Mean error of the apparent diameter and/or absolute radius
\item L Value in the catalogue obtained by a particular law of limb darkening
specified by the authors; 
Remarks give other values of the uniform disk and/or values corrected for the 
limb darkening. Details on this problem were outlined by Scholz (1997).
\item W Wavelength (or bandwidth) of measurement of apparent diameter and/or 
absolute radius and/or apparent magnitude when different from the visual one.
\end{enumerate} 
\end{enumerate}

\begin{table}
\begin{center}
\begin{tabular}{cccccccccc}
\hline
Method  &  1   &  2   &  3  &  4  & 5  & 6  & 7 & 8 & 9 \\
Stars   & 5805 & 1269 & 437 & 178 & 59 & 15 & 8 & 6 & 1 \\ 
\hline
\end{tabular}
\end{center}
\caption{Number of stars measured by one or more methodologies.}
\label{tab:2}
\end{table}

\section{Statistics and comments}

The statistics which follow will give a general view on the acquisition and the 
contents of the material gathered for this catalogue. The data added in 
the third edition concern 1062 stars; 
only 523 stars, however, are new entries. Considering that the first
edition reported 4266 stars, 6313 records, we can remark that 
the acquisition of data 
has developed faster in the second edition (1979-1985, 7255 stars, 12055
records) than in the third one (1986-1997).
The number of data reported with an estimate of the error increases from 9\% 
in the first edition to 11\% (IIed.) up to 15\% in the third edition. 
However, the error is given in about 45\% of the data added in the 
third edition ({\it new data}). 
Table\,\ref{tab:2} gives the number of stars measured by one or more
methodologies.

\begin{figure}
\begin{center}
\begin{tabular}{c}
\rotatebox{270}{\resizebox{6.3cm}{!}{\includegraphics{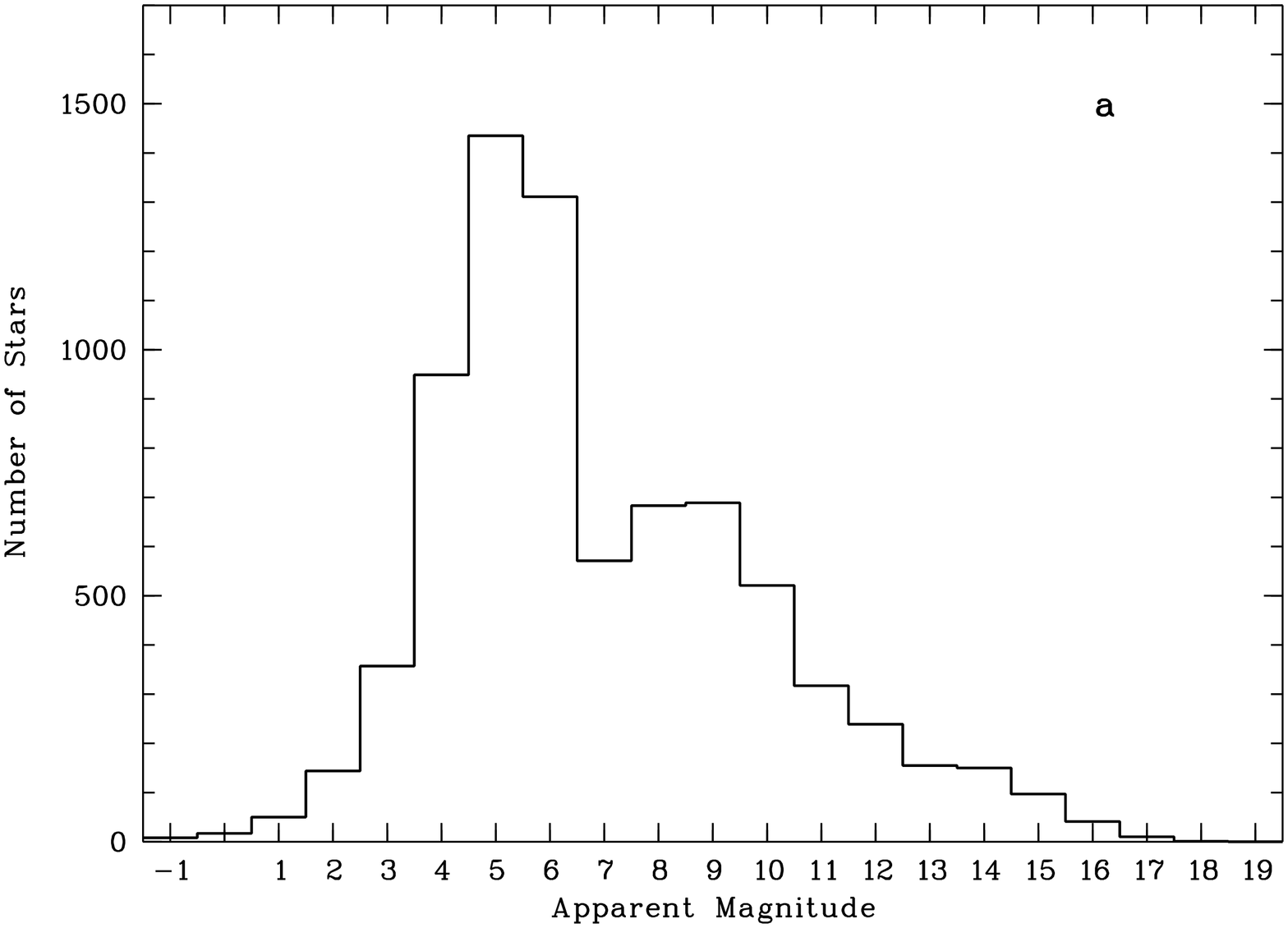}}} \\
\rotatebox{270}{\resizebox{6.3cm}{!}{\includegraphics{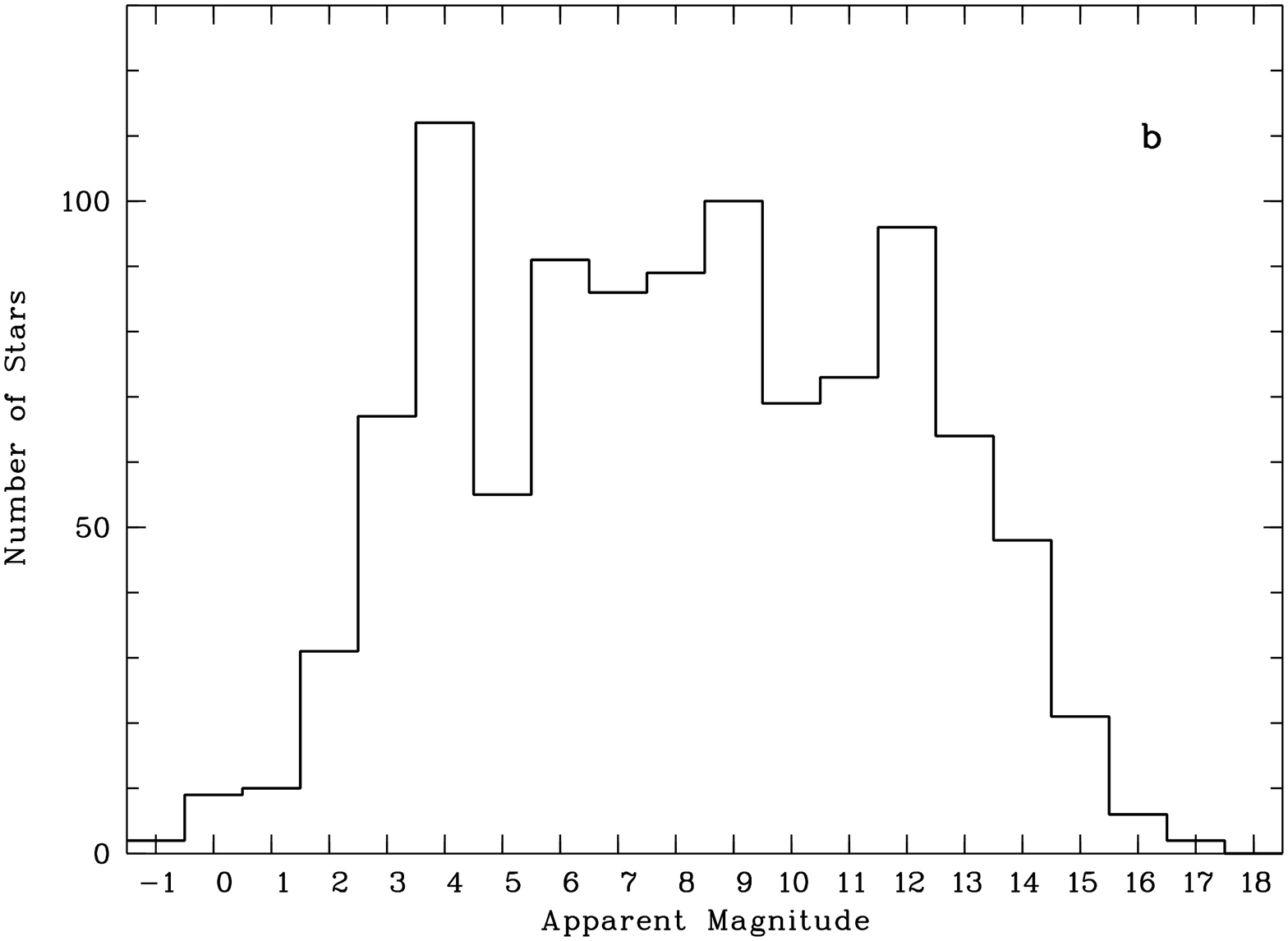}}} \\
\end{tabular}
\end{center}
\caption{Number of stars as a function of the apparent magnitude for
the whole sample of stars (panel {\bf a}) and for the {\it new data} (panel
{\bf b}).}
\label{fig:unica}
\end{figure}

The panels of Fig.\,\ref{fig:unica} give the number of stars as a 
function of the 
apparent magnitude for the whole sample of stars and for only the {\it new
data}. The magnitudes range from -1.5 down to 18.2. The comparison shows
significative differences; while 
the histogram of Fig.\,\ref{fig:unica}{\bf a} roughly reflects the natural 
frequency of the bright stars, influenced also by the difficulty for several 
methods to measure 
faint stars, the histogram of Fig.\,\ref{fig:unica}{\bf b} is more influenced by the scientific
interests of the authors and by the potentialities of new technologies.
Maxima correspond to 
magnitude 4, 9, 12 instead of 4, 5, 6. The measured stars with $m>12$ are 
increased up to 16\% in the {\it new data} while, considering the global 
number of data, they are about 6\% and 5\% in the third and second edition, 
respectively. Of course, the {\it new data} are also affected by the absence of 
determinations made by some methods (6F, which strongly affected the statistics 
of the previous editions, and 5). For analogous causes, M stars, white dwarfs 
and F stars are the most numerous among the {\it new data} instead of B, A, K 
types as in the whole sample (see also previous editions of CADARS). In 
particular, the maximum at the A-type reflects the natural frequency of
the stars, biased also by the high number of stars measured by the
method of Fracassini and Pasinetti (1967, 1973) around this spectral type.
The luminosity class is available in 76\% of the stars. Their distribution 
is as follows: I--II class 654 stars, III 1227, IV--V 4057.

\begin{table}
\begin{center}
\begin{tabular}{lcccccl}
\hline
Method &             & $\epsilon<1$\%  &  $\epsilon<5$\%  &  $\epsilon<10$\% 
&  min $\epsilon$ & Ref. \\
\hline
1          &   $\theta$  &   8.2    &     39.5  &    78.0   &   0.6   &   M21  
\\
2          &   $\theta$  &   3.5    &     23.6  &    47.1   &   0.1   &   Q3   
\\
3B         &   $\theta$  &   0.0    &     25.9  &    55.6   &   2.0   &   L4   
\\
4          &   r         &   8.4    &     63.5  &    81.4   &   0.1   &   C11  
\\
5          &   r         &   5.2    &     39.7  &    81.0   &   0.8   &   F9   
\\
5A         &   r         &   0.2    &     22.1  &    52.9   &   0.7   &   M17  
\\
5B         &   r         &   3.1    &     76.9  &    93.9   &   0.6   &   B20  
\\
6G         &   r         &   0.3    &      2.6  &    45.5   &   0.8   &   P30  
\\
6G         &   $\theta$  &  22.0    &     81.7  &    95.4   &   0.3   &   R13  
\\
6H         &   r         &   0.0    &      8.5  &    37.9   &   1.3   &   C34  
\\
6I         &   r         &   0.0    &      4.4  &    67.4   &   3.0   &   M23  
\\
6I         &   $\theta$  &   2.3    &     28.1  &    58.4   &   0.5   &   B13  
\\
6M         &   r         &  33.0    &     66.7  &    66.7   &   0.4   &   S36  
\\
\hline
\end{tabular}
\end{center}
\caption {Percentages of data with errors less than 1\%, 5\% and 10\%
for the different methods. (r): error on the absolute radius;
($\theta$): error on the apparent diameter. Last two columns:
minimum error of measure and related reference.}
\label{tab:3}
\end{table}

The methods 5, 6C, 6D, 6F, 6K, 6L were not utilized in the last decade. The new 
code 6M includes 14\% of the {\it new data}. The most utilized direct method 
is that of the eclipsing and spectroscopic binaries. The interferometric 
measurements are about doubled from the IInd 
to IIIrd edition (0.6\%, 1.2\% respectively) while in the {\it new data} their 
percentage is roughly comparable to that of binary stars (7.4\% and 9.3\% 
respectively).
The limits of magnitude and spectral type for each method (discussed by
Fracassini et al. 1981a, Pastori et al. 1988) are not significantly changed 
from the II to III edition.

\begin{table}
\begin{center}
\begin{tabular}{lcrrcl}
\hline
HD          &   met. &   II     &  III  &  Reference \\
\hline
29139      &    1   &    2.0   &  0.7  &  D10;Q2  \\ 
48915      &    1   &    2.3   &  1.4  &  H9;D16  \\
61421      &    1   &    6.8   &  1.0  &  H9;M21  \\
124897     &    1   &    5.7   &  1.0  &  C15;D15 \\
148478     &    1   &   12.1   &  2.0  &  W22;S34 \\
213306     &    5A  &    7.5   &  2.3  &  I7;T5   \\
102212    &    2   &    2.9   &  2.9  &  S17;S35 \\
223075    &    2   &    2.4   &  0.6  &  B2;R28  \\
34364G    &    4   &    5.6   &  2.0  &  C11;N10 \\
34364S    &    4   &    5.6   &  2.0  &  C11;N10 \\
40183G    &    4   &    3.2   &  1.1  &  K2;N9   \\
156247G   &    4   &    1.2   &  1.8  &  C11;H21 \\
156247S   &    4   &    3.3   &  1.7  &  C11;H21 \\
218066S   &    4   &    4.3   &  2.4  &  C11;C26 \\
\hline
\end{tabular}
\end{center}
\caption{Minimum errors determined in some stars by direct methods  
and related references.}
\label{tab:4}
\end{table}

Finally, we have considered the problem of the errors of determination which may 
be useful in the tests for new instruments or other scientific applications. 
Table\,\ref{tab:3} gives for the principal methods the percentage of measures 
reported with values of the errors less than 1\%, 5\%, 10\%. The last two 
columns give for each method the minimum error of measure reported in CADARS and 
the relative reference according to the used codification. As expected, if we 
consider only direct methods, methods 4, 1, 5, 2 (in order of percentage) are 
predominant for $\epsilon < 1$\%. Moreover, among all the data, the lowest value 
of error was obtained from the method 2 (diffraction--lunar occultation) 
followed by 4 (binaries) and 1 (interferometer). 
To test the improvement of the measures during the last decade, we have reported 
in Table\,\ref{tab:4} the minimum error of determination listed in the IInd and 
IIIrd editions for some stars reported by Pastori et al. (1985). With the 
exception of HD\,102212 and HD\,156247 the improvement is remarkable.

\section{Conclusions}

The quality of the measures reported in CADARS is noticeably improved as shown 
by the percentages of data obtained from direct methods, about 12\% and 15\% 
in the IInd and IIIrd edition respectively. This percentage increases up to 
41\% if we consider only the {\it new data}. Among the spectral types, the 
highest increase of measurements corresponds to the WDs (about 32\%), as it 
could be expected. Among the new data, the number of WDs is about equal to 
that of M stars reflecting the efficiency of new technologies. Considering the 
new data and those of the IIIrd and IInd edition, we find that M stars are about 
21\%, 16\%, 12\% respectively, and WDs 20\%, 5\%, 4\%. It is also 
increased the percentage of faint stars, while the improvement of the measures 
is shown by the percentage of data given with an estimate of the error and by 
the lowering of the error values.

\begin{acknowledgements} The authors are indebted to Dr. Francois Ochsenbein 
of the CDS for the careful tests on the CADARS. 
\end{acknowledgements}


\begin{thebibliography}{123456}
\bibitem{FP67}
Fracassini M., Pasinetti L.E., 1967, Atti XI Riunione SAIt, Padova
\bibitem{FGP73}
Fracassini M., Gilardoni G., Pasinetti L.E., 1973, Ap\&SS 22, 141
\bibitem{FP79}
Fracassini M., Pasinetti L.E., 1979, Bull. Inf. CDS n.16, 49
\bibitem{FMP81}
Fracassini M., Manzolini F., Pasinetti L.E., 1981a, A\&AS 44, 55
\bibitem{FPM81}
Fracassini M., Pasinetti L.E., Manzolini F., 1981b, A\&AS 45, 145
\bibitem{FPPP88}
Fracassini M., Pasinetti Fracassini L.E., Pastori L., Pironi R., 1988,
Bull.Inf.CDS, n.35, 121
\bibitem{PPAM85}
Pastori L., Pasinetti L.E., Antonello E., Malaspina G., 1985, IAU Symp. 111, 455
\bibitem{PPP88}
Pastori L., Pasinetti Fracassini L.E., Pironi R., 1988, Bull. Inf. CDS n.35, 69
\bibitem {S97}
Scholz M., 1997, IAU Symp. 189, 51
\bibitem{SIMBAD}
SIMBAD http://cdsweb.u-strasbg.fr/Simbad.html
\end{thebibliography}
\end{document}